\documentclass[runningheads]{llncs}

\usepackage{multirow}
\usepackage{graphicx}
\usepackage[table,xcdraw]{xcolor}
\usepackage{booktabs}
\usepackage{amsmath}
\usepackage{tikz}
\usepackage{hyperref}

\usepackage[normalem]{ulem}
\newcommand{\ftextbf}[1]{%
    \pdfliteral direct {2 Tr 0.3 w} 
     #1%
    \pdfliteral direct {0 Tr 0 w}%
}

\begin{document}
\title{Faster Diffusion Cardiac MRI with Deep Learning-based breath hold reduction\thanks{This work was supported in part by the UKRI CDT in AI for Healthcare \url{http://ai4health.io} (Grant No. EP/S023283/1), the British Heart Foundation (RG/19/1/34160), and the UKRI Future Leaders Fellowship (MR/V023799/1).}}

\titlerunning{Faster Diffusion Cardiac MRI with DL-based breath hold reduction}
%
%
\author{Michael T\"{a}nzer*\inst{1}\orcidID{0000-0002-9046-1008} \and
Pedro Ferreira*\inst{1,2}\orcidID{0000-0002-0436-3496} \and
Andrew Scott\inst{1,2}\orcidID{0000-0001-7656-3123} \and
Zohya Khalique\inst{1,2}\orcidID{0000-0002-0977-4796} \and
Maria Dwornik\inst{1,2}\orcidID{0000-0003-0843-8026} \and
Dudley Pennell\inst{1,2}\orcidID{0000-0001-5523-1314} \and
Guang Yang**\inst{1,2}\orcidID{0000-0001-7344-7733} \and
\\ Daniel Rueckert**\inst{1,3}\orcidID{0000-0002-5683-5889} \and
\\ Sonia Nielles-Vallespin**\inst{1,2}\orcidID{0000-0003-0412-2796}}
%
\authorrunning{M. T\"{a}nzer et al.}
%
\institute{Imperial College London \and
Royal Brompton and Harefield hospital \and
Technische Universität München (TUM) \\ *Send correspondence to \{m.tanzer,p.f.ferreira05,g.yang\}@imperial.ac.uk \\ **Co-last senior authors}
\maketitle
\begin{abstract}
    Diffusion Tensor Cardiac Magnetic Resonance (DT-CMR) enables us to probe the microstructural arrangement of cardiomyocytes within the myocardium in vivo and non-invasively, which no other imaging modality allows. This innovative technology could revolutionise the ability to perform cardiac clinical diagnosis, risk stratification, prognosis and therapy follow-up. However, DT-CMR is currently inefficient with over six minutes needed to acquire a single 2D static image. Therefore, DT-CMR is currently confined to research but not used clinically. We propose to reduce the number of repetitions needed to produce DT-CMR datasets and subsequently de-noise them, decreasing the acquisition time by a linear factor while maintaining acceptable image quality. Our proposed approach, based on Generative Adversarial Networks, Vision Transformers, and Ensemble Learning, performs significantly and considerably better than previous proposed approaches, bringing single breath-hold DT-CMR closer to reality.
    
\end{abstract}

\section{Introduction}
    Diffusion Tensor Cardiac Magnetic Resonance (DT-CMR) is the only medical imaging modality that allows us to non-invasively interrogate the micro-structure of the beating heart at a scale and resolution that other modalities cannot achieve \cite{moriPrinciplesDiffusionTensor2006}. In clinical research studies, DT-CMR has been shown to be useful in phenotyping several cardiomyopathies such as hypertrophic cardiomyopathy (HCM) and dilated cardiomyopathy (DCM) by quantitatively analysing the microstructural organisation and orientation of cardiomyocytes within the myocardium. DT-CMR also has the additional advantage of not requiring any contrast agent, which may be burdensome for patients with reduced kidney function \cite{schlaudeckerGadoliniumAssociatedNephrogenicSystemic2009}.
    
    In its current state, the acquisition time prevents clinical translation as around six minutes are needed to acquire a single 2D slice. For a typical acquisition protocol we require a minimum of three slices (basal, mid, apical), at least 7 different diffusion encoding steps and two time points of the cardiac cycle (systole and diastole), totalling 60 breath-holds and 90 minutes and making it clinically unfeasible. The long scan times have various source, but, most importantly, the protocol acquires multiple repetitions of each image to increase the signal-to-noise ratio (SNR) and to reduce motion-related artefacts. In this study, we will tackle the problem by reducing the number of repetitions.

\begin{figure}[hbt]
        \centering
        \includegraphics[width=1\textwidth]{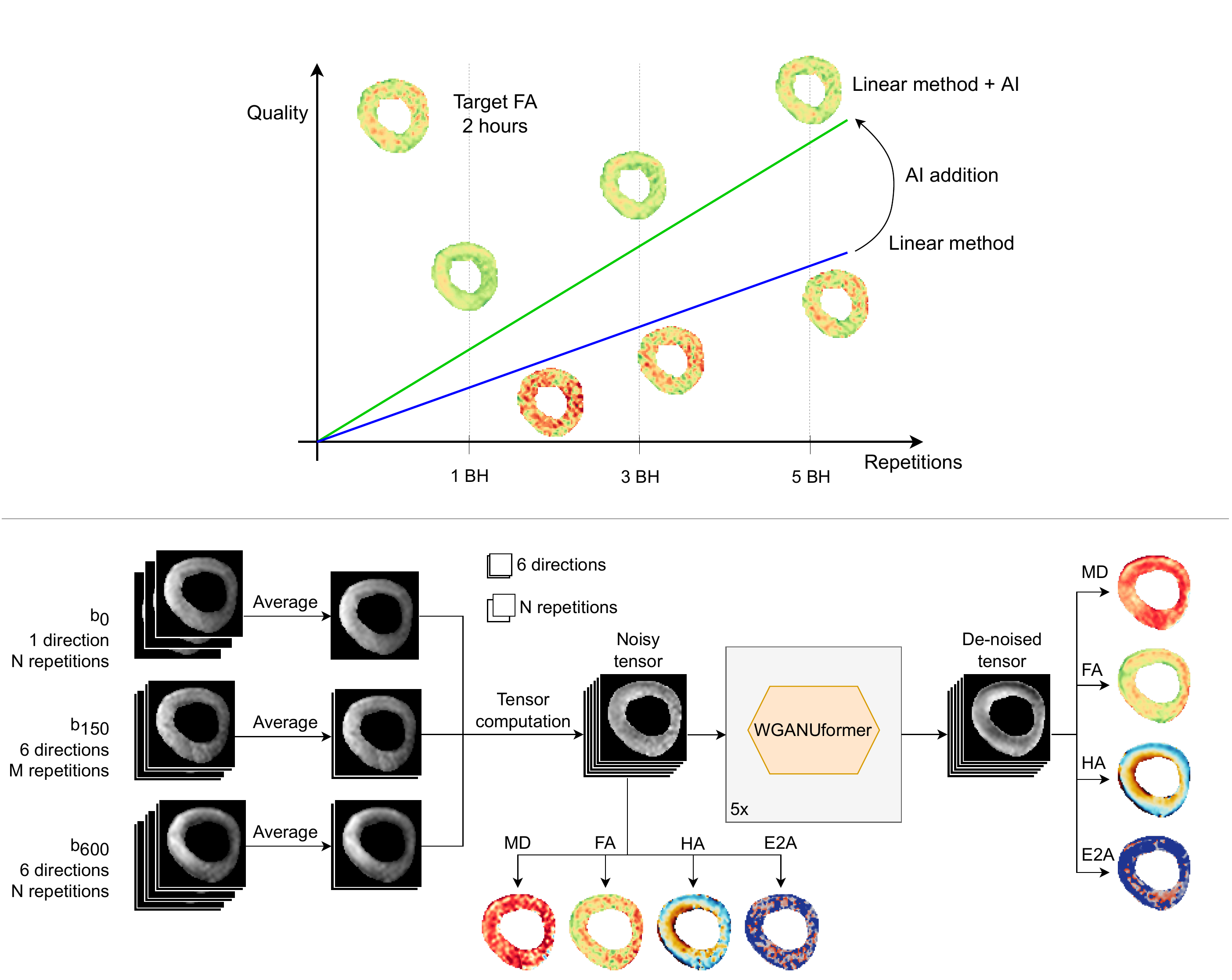}
        \caption{Proposed deep learning framework. From left to right, we can see the original input data comprised of several repetitions that are then averaged to increase the SNR and reduce artefacts. This averaged data is then used to compute (noisy) diffusion tensors using a least squares tensor fit. From the noisy tensors we can then compute noisy DT-CMR maps (lower central part of the image). In our proposed framework we use an ensemble of deep-learning models to de-noise the diffusion tensors and therefore obtain better DT-CMR maps (shown on the right).}
        \label{fig:arch_images}
    \end{figure}

    Our contribution, shown in Figure \ref{fig:arch_images}, is a novel deep-learning framework that can be used to reduce the number of repetitions used in the DT-CMR acquisition. Using fewer averages leads to lower SNR, and deep learning can be used to recover the original full-repetition data. This allows us to greatly reduce the total acquisition time with minimal loss of image quality. This method could potentially be adopted to acquire a DT-CMR scan in only one breath hold while maintaining acceptable quality, reducing the scan time from several minutes to well under a minute.
    
\section{Background}\label{sec:bg}
\subsection{DT-CMR}\label{sec:bg:dtcmr}
    DT-CMR measures the diffusion pattern of water molecules in every voxel of the imaged tissue and approximates it with a 3D tensor. As the free diffusion of water in the tissue is constrained by the shape of cardiac muscle microstructure for every voxel, studying the extracted 3D tensors has been shown to give us information related to the shape and orientation of the cardiomyocytes in the imaged tissue.
    
    In-vivo DT-CMR requires the rapid acquisition of multiple single-shot diffusion-weighted images with diffusion encoded in at least six different 3D directions. Single-shot encoding acquisitions translate to low SNR images. Low SNR is also an inherent issue in DT-CMR as we measure the signal lost due to diffusion. Therefore, multiple repetitions are commonly acquired to increase the quality of the signal. While improving the SNR, this also translates to longer acquisition times and extra breath-holds for the patient. Our clinical research protocol requires approximately 12 breath-holds for every single DT-CMR slice at a single time point in the cardiac cycle. The series of signal intensities is then fitted to a rank-2 diffusion tensor using a linear least-square (LLS) \cite{IntroductionDiffusionTensor} fitting or alternatively more advanced linear and non-linear iterative methods \cite{IterativeReweightedLinear}.
    
    The cardiac diffusion tensor information is commonly visualised and quantified through four per-voxel metric maps: Mean Diffusivity (MD) that quantifies the total diffusion in the voxel (higher corresponds to more diffusion), Fractional Anisotropy (FA) that quantifies the level of organisation of the tissue (higher corresponds to a higher organisation), Helix Angle (HA) and Second Eigenvector (E2) Angle (E2A) that quantify the 3D orientation and shape of the tissue in the voxel \cite{basser1995inferring,kung2011presence}.
    

\subsection{De-noising in DT-CMR}\label{sec:bg:denoising}
    There are several ways to approach the task of reducing the number of repetitions used to compute the DT-CMR maps. In our proposed approach, we will see how we treat it as de-noising task where the goal is to produce de-noised diffusion tensors from noisy tensors.

    De-noising is the process of removing noise from a given signal with the aim of restoring the original noise-free version of the signal. In a number of studies the noise is assumed to come from a known distribution, giving rise to models that work on this assumption to remove it \cite{zhangGaussianDenoiserResidual2017}, while some other studies instead do not make any assumption on the source of the noise and produce models that are more robust to real-world noise \cite{batsonNoise2SelfBlindDenoising2019}. In recent years, deep learning-based de-noising has been extremely popular, both applied to photographs \cite{xieImageDenoisingInpainting2012} and to other types of signals, such as MRI data \cite{jiangDenoising3DMagnetic2018,manjonMRIDenoisingUsing2018}, CT data \cite{chenLowdoseCTDenoising2017}, audio data \cite{xieImageDenoisingInpainting2012}, and point clouds \cite{hermosillaTotalDenoisingUnsupervised2019}. Focusing on de-noising models designed for imaging data, Batson et al. \cite{batsonNoise2SelfBlindDenoising2019} used a U-Net model and a self-supervised approach to blindly de-noise images without using an assumed noise distribution. Park et al. \cite{parkUnpairedImageDenoising2019} trained a de-noising model on unpaired CT data using a GAN model. More recently, Vision Transformers have also been used to tackle the de-noising problem \cite{liangSwinIRImageRestoration2021,zamirRestormerEfficientTransformer2021,chenPreTrainedImageProcessing2021,wangUformerGeneralUShaped2021}.
    
    Phipps et al. \cite{phippsAcceleratedVivoCardiac2021} used a residual-learning approach to de-noise the diffusion weighted images prior to the tensor calculation to reduce the number of acquisitions required to produce high-quality DT-CMR maps. In our previous work \cite{OURPAPERNOTPUBLIC}, our group also showed how a U-NET-based model can be successfully used to predict de-noised tensors directly from noisy images.

\section{Methods}
    The study is divided into two main sections: (1) the analysis of how the number and choice of repetitions affect the quality of the DT-CMR maps and (2) our proposed deep-learning-based de-noising procedure and the validation of its results.

\subsection{Data acquisition}
    All data used in this work was approved by the National Research Ethics Service. All subjects gave written informed consent.
    
    All the data were acquired using a Siemens Skyra 3T MRI scanner and more recently a Siemens Vida 3T MRI scanner (Siemens AG, Erlangen, Germany) with diffusion weighted stimulated echo acquisition mode (STEAM) single shot echo planar imaging (EPI) sequence with reduced phase field-of-view and fat saturation, TR = 2RR intervals, TE = 23 ms, SENSE or GRAPPA R = 2, echo train duration = 13 ms, at a spatial resolution of 2.8 x 2.8 x 8.0 mm$^3$. Diffusion was encoded in six directions with diffusion- weightings of b = 150 and 600 s/mm$^2$ in a short-axis mid-ventricular slice. Additionally, reference images were also acquired with minimal diffusion weighting, named here as ``b$_0$" images. All diffusion data were acquired under multiple breath-holds, each with a duration of 18 heartbeats.  
    We used a total of 744 DT-CMR datasets, containing a mixture of healthy volunteers (26\%, n = 197) and patient (74\%, n = 547) scans acquired in either the diastolic pause (49\%, n = 368) or end-systole (51\%, n = 376). The patient data comes from several conditions including 31 amyloidoses, 45 dilated cardiomyopathy (DCM), 11 Fabry’s disease, 48 HCM genotype-positive-phenotype-negative (HCM G+P-), 66 hypertrophic cardiomyopathies (HCM), 4 hypertensive DCM (hDCM), 246 acute myocardial infarction (MI), 7 Marfan’s syndrome, and 89 in-recovery DCM (rDCM).  


\subsection{Data preparation}
    The mean number of repetitions was 12 ± 2.0 for b$_0$ images; 10 ± 2.2 for b = 600 s/mm$^2$ images; and 2 ± 0.6 for b = 150 s/mm$^2$. These datasets, containing all acquired data, were used to calculate the reference tensor results for each subject using a newly developed tool written in Python and validated against our previous post-processing software
    \cite{OURPAPERNOTPUBLIC2}
    . Before tensor calculation, all the diffusion images were assessed visually, and images corrupted with bulk motion artefacts were removed. Subsequently, all remaining images were registered with a multi-resolution rigid sub-pixel translation algorithm \cite{guizar-sicairosEfficientSubpixelImage2008}, manually thresholded to remove background features. Lastly, the left ventricle (LV) myocardium was segmented excluding papillary muscle.  
    
    Tensors were calculated with an LLS fit of all the acquired repetitions and respective diffusion weightings and directions \cite{IntroductionDiffusionTensor}. The tensors were then used to compute the DT-CMR maps that we considered as the ground truth for all the comparisons in the study.  
    
    We were also able to dynamically create three new datasets with an increasingly reduced number of repetitions. We assessed the quality or the DT-CMR maps produced from different subsets of repetitions (e.g., using the first N repetitions vs using the last N repetitions, see Section \ref{sec:methods:effect-reps}). We proposed three choices for the numbers of repetitions that result in three datasets:
    \begin{itemize}
        \item 5BH. Four repetitions of b$_0$ and b$_{600}$, and one repetition of b$_{150}$. This acquisition would require 5 breath-holds.
        \item 3BH. Two repetitions of b$_0$ and $b_{600}$ , and one repetition of b$_{150}$. This acquisition would require 3 breath-holds.
        \item 1BH. One repetition of b$_0$ and b$_{600}$ only. This acquisition would require 1 breath-hold.
    \end{itemize}
    
    
    For the purpose of training a deep-learning model, the data was also randomly augmented with random rotation and random cropping.

\subsection{The effect of repetitions}\label{sec:methods:effect-reps}
    In a standard DT-CMR acquisition, we acquire several repetitions to reduce the effect of noise and motion. To do so, we ask the patient to hold and resume their breathing at fixed intervals.
    
    First, we quantitatively studied how the number of repetitions (and breath-holds) affects the quality of the maps. We compared the maps produced from all available breath-holds with maps computed from $M$ repetitions where $M$ represents a number smaller than the number of available repetitions for the patient. We repeated the process separately for the four maps.
    
    Secondly, we analysed how choosing different repetition subsets affects the final quality of the DT-CMR acquisition. We defined five different methods to select a subset from the eight original repetitions: (1) we selected the first $M$ repetitions (First, F); (2) we selected the central $M$ repetitions (Centre, C); (3) we used the last $M$ repetitions (Last, L); (4) uniformly random repetition sampling (Random, R); (5) based on the clinician's observation that the first breath-hold is usually lower-quality due to the patient adjusting to holding and resuming their breath, we selected the first $M+1$ repetitions and discarded the first one (First+1, F1).

\subsection{Deep-learning-based de-noising}
    In this study, we developed and trained a deep-learning model based on the current state-of-the-art architectures for image de-noising to improve the quality of noisy DT-CMR tensors produced from the low-breath-holds datasets described above.
    
    \subsubsection{Input and output}
    The output of all the models reported here is the diffusion tensor components. As the tensors are represented by a rank-2 symmetric matrix, they only contain six unique components: for a $3 \times 3$ matrix D we only need the upper triangular elements to represent it. All input and output images were cropped to be $128 \times 128$ pixels in size. Thus, the output is a $128 \times 128$ image with 6 channels. For the input we compared different approaches: building on our previous work, we used the average diffusion weighted images as input, resulting in 13 channels for the 5BH and 3BH datasets (1 b$_0$ + 6 b$_{600}$ + 6 b$_{150}$) and 7 channels for the 1BH dataset (1 b$_0$ + 6 b$_{600}$). Alternatively, we also experimented with de-noising diffusion tensors directly, which translated to a six-channel input image.

    We had two types of inputs: diffusion-weighted images (DWI) or diffusion tensors. In the case of DWI, the images were normalised in the range $[0, 1]$ by dividing them by the maximum value present in the dataset. The background pixels were also replaced with zeros. The diffusion tensors were instead either normalised by a fixed amount (500) that was empirically found to make most values in the range $[-1, 1]$ or normalised by performing channel-wise z-score normalisation across the whole dataset. In the latter case, the normalisation was undone before computing the maps.
    
    The data were randomly divided into three parts: a training set, a validation set, and a test set with ratios of 80:10:10 respectively. In order to ensure consistency, all the experiments maintained the same random split.
    
   \subsubsection{Model}\label{sec:methods:models}
   We compared our new model with our previous work as the setting is extremely similar to the one proposed here and we had obtained promising results. In our previous setting, we used a U-Net model with six encoders and six decoders. Each encoding layer consisted of two blocks, each containing a convolution layer, a batch normalisation layer and the leaky ReLU activation function; after the two blocks, a max-pooling operator was applied. Each decoding layer consisted of two blocks: the first one contained a transpose convolution, a concatenation operation, batch normalisation, and the leaky ReLU activation; the second one instead consisted of a convolution, batch normalisation, and the leaky ReLU activation. The concatenation was between encoding and decoding layers as per the original U-Net formulation. This baseline model contained a total of 31 million trainable parameters.
   
   We proposed several modifications to the baseline above. To show the effect of these changes, we progressively introduced them on the baseline model to show how they affected the quality of the output. In these experiments, we kept the main structure of the model unchanged and we did not modify the training hyperparameters. Specifically, in order, we experimented with:
   \begin{enumerate}
       \item U-NET image-to-tensor baseline model.
       \item Baseline with channel normalisation (BL+CN): the output tensors were normalised with a channel-wise z-score normalisation.
       \item Baseline with tensor-to-tensor training (BL+T2T): the input type was changed from images to tensors, making the task a tensor de-noising task. This also allowed us to use residual learning for our training, improving convergence and performance. For this experiment, the tensors were only normalised by dividing all the values by a fixed amount.
       \item BL+CN with tensor to tensor (BL+CN+T2T): similarly to BL+T2T, the training was performed on a tensor-to-tensor de-noising task, but in this case, the input and target tensors were normalised with z-score normalisation.
       \item BL+CN+T2T with multiple datasets (BL+CN+multiT2T): multiple repetition strategies were used simultaneously for the training (First, Centre, Last). This made significantly better use of the available training data and effectively increased the size of the dataset by a factor of 3 (although using non-independent data for the training).
   \end{enumerate}
   
    State of the art (SOTA) models in image de-noising and image restoration were also investigated: Restormer \cite{zamirRestormerEfficientTransformer2021} and Uformer \cite{wangUformerGeneralUShaped2021}. These models were trained with Channel Normalisation and multi tensor-to-tensor.
    
    Finally, we investigated a novel model that made use of all the additions proposed above based on SOTA models. Specifically, we expanded on the Uformer architecture by using it as the generator of a generative adversarial network. The Uformer is a U-Net-like Transformer-based architecture that uses LeWin blocks and residual connections to form a hierarchical encoder-decoder network. 
    
    We, therefore, proposed the WGANUformer (WGUF) by adding a PatchGAN discriminator \cite{demirPatchBasedImageInpainting2018} and an adversarial loss to a Uformer. The model was trained with a Wasserstein objective function and with weight clipping as per Arjovsky et al. \cite{arjovskyWassersteinGAN2017} as it combats mode collapse and has been proved to converge to optimality, unlike other GAN formulations.
    
    A schematic representation of the architecture and training procedure can be found in Figure \ref{fig:arch_images}.
    
    
    \subsubsection{Training}
    All training was performed using a workstation with Ubuntu 18.04, CPU Intel i7-10700k, 64 GB of RAM, and an NVIDIA RTX 3080 GPU (Python 3.8 with PyTorch 1.9). The training required a total of 215 GPU-hours, resulting in an estimated 30 kg CO$_2$eq.
    
    During training, a mean absolute error was used as the loss function. For the baseline-based models, other parameters included an Adam optimiser with a learning rate = $10^{-4}$, beta1 = 0.9, beta2 = 0.999; a batch-size of 8 images and 500 epochs. These parameters and CNN design were optimised empirically based on our pilot study results. The WGANUformer was trained as per Loshchilov et al. with the AdamW optimiser \cite{loshchilovDecoupledWeightDecay2019}, a learning rate = $10^{-4}$, beta1 = 0.9, beta2 = 0.999, weight decay alpha = 0, a batch-size of 8 tensors for 500 epochs. All models were trained from scratch without any pre-training.
    
    In all our experiments, we report the metrics computed on the never-seen-before test set using the model that produced the lowest validation loss.

\subsection{DT-CMR post-processing}  
    For the computation of all metrics and maps, we post-processed the data with an in-house developed software written in Python. For the post-processing, each subject in the dataset was processed several times:
    \begin{itemize}
        \item Initially to obtain the reference tensor parameter results using all available repetitions.
        \item Every time we computed a dataset with a reduced number of repetitions we performed the same steps (image registration, thresholding, and segmentation). When training the models using images as input, we replaced the tensor calculation process with the model prediction.
        \item To produce a comparison, we also computed the conventional LLS tensor fit from the reduced datasets. All the comparisons are voxel-wise.
    \end{itemize}

\subsection{DT-CMR maps comparison}
    Four DT-CMR maps representing different aspects of the diffusion of water within the tissue were chosen as output. These maps represent different physical properties and need to be compared with appropriate metrics.

    HA and E2A are angular maps with values between -90$^{\circ}$ and 90$^{\circ}$. When comparing these maps, we were interested in the direction of the vector corresponding to the angle but not its orientation. This means that any two angles with a 180$^{\circ}$ difference should be identical and two angles with a 90$^{\circ}$ difference should have the maximum distance. For HA and E2A we then reported the Mean Angle Absolute Error (MAAE):
    \begin{equation}
        \text{MAAE}(X, Y) = \frac{1}{NM} \sum_{i=0}^{N} \sum_{j=0}^{M} \begin{cases}
            \left|X^{(i,j)} - Y^{(i,j)}\right|,\ \text{if}\ \left|X^{(i,j)} - Y^{(i,j)}\right| < 90^\circ\\
            180^\circ - \left|X^{(i,j)} - Y^{(i,j)}\right|,\ \text{otherwise}
        \end{cases}
    \end{equation}
    
    For MD and FA, as they are scalar maps, we therefore reported the Mean Absolute Error (MAE) between de-noised and target maps.
    
    In the experiments below, we reported the MAAE and MAE across all the voxels in the left ventricle (i.e., ignoring the background and the right ventricle).
    
    

\subsubsection{Statistical analysis}
    We treated all results as non-parametric as we were unable to assure normal distributions in the test subjects. The statistical significance threshold for all tests was set at P = 0.05. Intersubject measures are quoted as \textit{median [interquartile range]}.

\section{Results}

\subsection{The effect of repetitions}
    From Table \ref{tab:bh-sampling-patterns}, no significant differences were found between the distribution of errors for HA, E2A, and FA for all the pairs of strategies only containing First, Centre, First+1, and Last. There is, instead, a significant difference (P$<$0.05) in distributions between the Random strategy compared to the other strategies for these metrics (with few non-significant exceptions for FA in 3BH and 5BH). When analysing the MD MAE errors, we found that the pairwise significance pattern we had observed for the other metrics does not hold and we did not recognise any clear pattern.
    

\begin{table}[!htb]
\centering
\footnotesize
    \begin{tabular}{@{}c|l|rcccc|rcccc@{}}
    \toprule
    \textbf{Dataset} & \textbf{Scheme} & \multicolumn{5}{c|}{\textbf{HA MAAE}} & \multicolumn{5}{c|}{\textbf{E2A MAAE}} \\ \specialrule{.1em}{.1em}{.1em}

    \multirow{5}{*}{1BH} & First & 25.69 {[}4.95{] } & \cellcolor[HTML]{F4C7C3}C & \cellcolor[HTML]{F4C7C3}L & \cellcolor[HTML]{F4C7C3}F1 & \cellcolor[HTML]{B7E1CD}R  & 33.66 {[}4.77{] }     & \cellcolor[HTML]{F4C7C3}C & \cellcolor[HTML]{F4C7C3}L & \cellcolor[HTML]{F4C7C3}F1 & \cellcolor[HTML]{B7E1CD}R \\
    
    &Centre & 25.93 {[}5.26{] } & \cellcolor[HTML]{F4C7C3}F & \cellcolor[HTML]{F4C7C3}L & \cellcolor[HTML]{F4C7C3}F1 & \cellcolor[HTML]{B7E1CD}R  & 33.86 {[}4.57{] }     & \cellcolor[HTML]{F4C7C3}F & \cellcolor[HTML]{F4C7C3}L & \cellcolor[HTML]{F4C7C3}F1 & \cellcolor[HTML]{B7E1CD}R \\
    
    &Last & 25.94 {[}4.91{] } & \cellcolor[HTML]{F4C7C3}F & \cellcolor[HTML]{F4C7C3}C & \cellcolor[HTML]{F4C7C3}F1 & \cellcolor[HTML]{B7E1CD}R  & 33.87 {[}4.33{] }     & \cellcolor[HTML]{F4C7C3}F & \cellcolor[HTML]{F4C7C3}C & \cellcolor[HTML]{F4C7C3}F1 & \cellcolor[HTML]{B7E1CD}R \\
    
    &First+1 & 25.69 {[}4.82{] } & \cellcolor[HTML]{F4C7C3}F & \cellcolor[HTML]{F4C7C3}C & \cellcolor[HTML]{F4C7C3}L  & \cellcolor[HTML]{B7E1CD}R  & 33.57 {[}4.70{] }     & \cellcolor[HTML]{F4C7C3}F & \cellcolor[HTML]{F4C7C3}C & \cellcolor[HTML]{F4C7C3}L  & \cellcolor[HTML]{B7E1CD}R \\
    
    &Random & 26.60 {[}4.78{] } & \cellcolor[HTML]{B7E1CD}F & \cellcolor[HTML]{B7E1CD}C & \cellcolor[HTML]{B7E1CD}L  & \cellcolor[HTML]{B7E1CD}F1 & 34.36 {[}4.48{] }     & \cellcolor[HTML]{B7E1CD}F & \cellcolor[HTML]{B7E1CD}C & \cellcolor[HTML]{B7E1CD}L  & \cellcolor[HTML]{B7E1CD}F1 \\\specialrule{.1em}{.1em}{.1em}
    
    \multirow{5}{*}{3BH} & First & 20.10 {[}4.62{] } & \cellcolor[HTML]{F4C7C3}C & \cellcolor[HTML]{F4C7C3}L & \cellcolor[HTML]{F4C7C3}F1 & \cellcolor[HTML]{B7E1CD}R  & 28.13 {[}5.45{] }     & \cellcolor[HTML]{F4C7C3}C & \cellcolor[HTML]{F4C7C3}L & \cellcolor[HTML]{F4C7C3}F1 & \cellcolor[HTML]{B7E1CD}R \\
    
    &Centre & 20.12 {[}4.75{] } & \cellcolor[HTML]{F4C7C3}F & \cellcolor[HTML]{F4C7C3}L & \cellcolor[HTML]{F4C7C3}F1 & \cellcolor[HTML]{B7E1CD}R  & 28.24 {[}5.04{] }     & \cellcolor[HTML]{F4C7C3}F & \cellcolor[HTML]{F4C7C3}L & \cellcolor[HTML]{F4C7C3}F1 & \cellcolor[HTML]{B7E1CD}R  \\
    
    &Last & 20.23 {[}4.89{] } & \cellcolor[HTML]{F4C7C3}F & \cellcolor[HTML]{F4C7C3}C & \cellcolor[HTML]{F4C7C3}F1 & \cellcolor[HTML]{B7E1CD}R  & 28.22 {[}4.94{] }     & \cellcolor[HTML]{F4C7C3}F & \cellcolor[HTML]{F4C7C3}C & \cellcolor[HTML]{F4C7C3}F1 & \cellcolor[HTML]{B7E1CD}R  \\
    
    &First+1 & 20.09 {[}4.63{] } & \cellcolor[HTML]{F4C7C3}F & \cellcolor[HTML]{F4C7C3}C & \cellcolor[HTML]{F4C7C3}L  & \cellcolor[HTML]{B7E1CD}R  & 28.23 {[}5.33{] }     & \cellcolor[HTML]{F4C7C3}F & \cellcolor[HTML]{F4C7C3}C & \cellcolor[HTML]{F4C7C3}L  & \cellcolor[HTML]{B7E1CD}R  \\
    
    &Random & 21.02 {[}4.74{] } & \cellcolor[HTML]{B7E1CD}F & \cellcolor[HTML]{B7E1CD}C & \cellcolor[HTML]{B7E1CD}L  & \cellcolor[HTML]{B7E1CD}F1 & 29.03 {[}5.07{] }     & \cellcolor[HTML]{B7E1CD}F & \cellcolor[HTML]{B7E1CD}C & \cellcolor[HTML]{B7E1CD}L  & \cellcolor[HTML]{B7E1CD}F1 \\\specialrule{.1em}{.1em}{.1em}

    \multirow{5}{*}{5BH} & First & 13.90 {[}4.13{] } & \cellcolor[HTML]{F4C7C3}C & \cellcolor[HTML]{F4C7C3}L & \cellcolor[HTML]{F4C7C3}F1 & \cellcolor[HTML]{B7E1CD}R  & 20.98 {[}5.28{] }     & \cellcolor[HTML]{F4C7C3}C & \cellcolor[HTML]{F4C7C3}L & \cellcolor[HTML]{F4C7C3}F1 & \cellcolor[HTML]{B7E1CD}R \\
    
    &Centre & 13.97 {[}4.08{] } & \cellcolor[HTML]{F4C7C3}F & \cellcolor[HTML]{F4C7C3}L & \cellcolor[HTML]{F4C7C3}F1 & \cellcolor[HTML]{B7E1CD}R  & 21.02 {[}5.30{] }     & \cellcolor[HTML]{F4C7C3}F & \cellcolor[HTML]{F4C7C3}L & \cellcolor[HTML]{F4C7C3}F1 & \cellcolor[HTML]{B7E1CD}R \\
    
    &Last & 14.07 {[}4.05{] } & \cellcolor[HTML]{F4C7C3}F & \cellcolor[HTML]{F4C7C3}C & \cellcolor[HTML]{F4C7C3}F1 & \cellcolor[HTML]{B7E1CD}R  & 20.89 {[}5.08{] }     & \cellcolor[HTML]{F4C7C3}F & \cellcolor[HTML]{F4C7C3}C & \cellcolor[HTML]{F4C7C3}F1 & \cellcolor[HTML]{B7E1CD}R \\
    
    &First+1 & 14.06 {[}4.00{] } & \cellcolor[HTML]{F4C7C3}F & \cellcolor[HTML]{F4C7C3}C & \cellcolor[HTML]{F4C7C3}L  & \cellcolor[HTML]{B7E1CD}R  & 21.07 {[}5.27{] }     & \cellcolor[HTML]{F4C7C3}F & \cellcolor[HTML]{F4C7C3}C & \cellcolor[HTML]{F4C7C3}L  & \cellcolor[HTML]{B7E1CD}R \\
    
    &Random & 14.68 {[}4.26{] } & \cellcolor[HTML]{B7E1CD}F & \cellcolor[HTML]{B7E1CD}C & \cellcolor[HTML]{B7E1CD}L  & \cellcolor[HTML]{B7E1CD}F1 & 21.78 {[}5.51{] }     & \cellcolor[HTML]{B7E1CD}F & \cellcolor[HTML]{B7E1CD}C & \cellcolor[HTML]{B7E1CD}L  & \cellcolor[HTML]{B7E1CD}F1 \\\specialrule{.1em}{.1em}{.1em}
    \multicolumn{5}{c}{ }\\\specialrule{.1em}{.1em}{.1em}
    
    \textbf{Dataset} & \textbf{Scheme} & \multicolumn{5}{c|}{\textbf{MD MAE ($\times 10^5$)}} & \multicolumn{5}{c|}{\textbf{FA MAE ($\times 10^2$)}}\\ \specialrule{.1em}{.1em}{.1em}

    \multirow{5}{*}{1BH} & First & 14.69 {[}5.51{] } & \cellcolor[HTML]{B7E1CD}C & \cellcolor[HTML]{B7E1CD}L & \cellcolor[HTML]{F4C7C3}F1 & \cellcolor[HTML]{B7E1CD}R & 19.50 {[}4.59{] } & \cellcolor[HTML]{F4C7C3}C & \cellcolor[HTML]{F4C7C3}L & \cellcolor[HTML]{F4C7C3}F1 & \cellcolor[HTML]{B7E1CD}R \\
    
    &Centre  & 15.75 {[}6.39{] } & \cellcolor[HTML]{B7E1CD}F & \cellcolor[HTML]{F4C7C3}L & \cellcolor[HTML]{B7E1CD}F1 & \cellcolor[HTML]{F4C7C3}R  & 19.80 {[}4.72{] } & \cellcolor[HTML]{F4C7C3}F & \cellcolor[HTML]{F4C7C3}L & \cellcolor[HTML]{F4C7C3}F1 & \cellcolor[HTML]{B7E1CD}R  \\
    
    &Last  & 15.58 {[}6.52{] } & \cellcolor[HTML]{B7E1CD}F & \cellcolor[HTML]{F4C7C3}C & \cellcolor[HTML]{B7E1CD}F1 & \cellcolor[HTML]{F4C7C3}R & 19.45 {[}4.82{] } & \cellcolor[HTML]{F4C7C3}F & \cellcolor[HTML]{F4C7C3}C & \cellcolor[HTML]{F4C7C3}F1 & \cellcolor[HTML]{B7E1CD}R  \\
    
    &First+1  & 14.44 {[}5.30{] } & \cellcolor[HTML]{F4C7C3}F & \cellcolor[HTML]{B7E1CD}C & \cellcolor[HTML]{B7E1CD}L  & \cellcolor[HTML]{B7E1CD}R & 19.50 {[}4.57{] } & \cellcolor[HTML]{F4C7C3}F & \cellcolor[HTML]{F4C7C3}C & \cellcolor[HTML]{F4C7C3}L  & \cellcolor[HTML]{B7E1CD}R  \\
    
    &Random  & 15.72 {[}6.52{] } & \cellcolor[HTML]{B7E1CD}F & \cellcolor[HTML]{F4C7C3}C & \cellcolor[HTML]{F4C7C3}L  & \cellcolor[HTML]{B7E1CD}F1 & 20.41 {[}5.01{] } & \cellcolor[HTML]{B7E1CD}F & \cellcolor[HTML]{B7E1CD}C & \cellcolor[HTML]{B7E1CD}L  & \cellcolor[HTML]{B7E1CD}F1 \\\specialrule{.1em}{.1em}{.1em}

    \multirow{5}{*}{3BH} & First & 9.76 {[}3.93{] } & \cellcolor[HTML]{F4C7C3}C & \cellcolor[HTML]{F4C7C3}L & \cellcolor[HTML]{F4C7C3}F1 & \cellcolor[HTML]{B7E1CD}R & 14.13 {[}3.92{] } & \cellcolor[HTML]{F4C7C3}C & \cellcolor[HTML]{F4C7C3}L & \cellcolor[HTML]{F4C7C3}F1 & \cellcolor[HTML]{F4C7C3}R  \\
    
    &Centre & 9.63 {[}3.76{] } & \cellcolor[HTML]{F4C7C3}F & \cellcolor[HTML]{B7E1CD}L & \cellcolor[HTML]{F4C7C3}F1 & \cellcolor[HTML]{B7E1CD}R & 13.99 {[}4.04{] } & \cellcolor[HTML]{F4C7C3}F & \cellcolor[HTML]{F4C7C3}L & \cellcolor[HTML]{F4C7C3}F1 & \cellcolor[HTML]{B7E1CD}R    \\
    
    &Last & 10.12 {[}4.36{] } & \cellcolor[HTML]{F4C7C3}F & \cellcolor[HTML]{B7E1CD}C & \cellcolor[HTML]{B7E1CD}F1 & \cellcolor[HTML]{B7E1CD}R & 14.03 {[}3.72{] } & \cellcolor[HTML]{F4C7C3}F & \cellcolor[HTML]{F4C7C3}C & \cellcolor[HTML]{F4C7C3}F1 & \cellcolor[HTML]{B7E1CD}R   \\
    
    &First+1  & 9.64 {[}3.91{] } & \cellcolor[HTML]{F4C7C3}F & \cellcolor[HTML]{F4C7C3}C & \cellcolor[HTML]{B7E1CD}L  & \cellcolor[HTML]{B7E1CD}R & 14.05 {[}3.92{] } & \cellcolor[HTML]{F4C7C3}F & \cellcolor[HTML]{F4C7C3}C & \cellcolor[HTML]{F4C7C3}L  & \cellcolor[HTML]{F4C7C3}R  \\
    
    &Random & 8.85 {[}3.38{] } & \cellcolor[HTML]{B7E1CD}F & \cellcolor[HTML]{B7E1CD}C & \cellcolor[HTML]{B7E1CD}L  & \cellcolor[HTML]{B7E1CD}F1 & 14.49 {[}4.05{] } & \cellcolor[HTML]{F4C7C3}F & \cellcolor[HTML]{B7E1CD}C & \cellcolor[HTML]{B7E1CD}L  & \cellcolor[HTML]{F4C7C3}F1 \\\specialrule{.1em}{.1em}{.1em}

    \multirow{5}{*}{5BH} & First & 6.24 {[}2.68{] } & \cellcolor[HTML]{F4C7C3}C & \cellcolor[HTML]{F4C7C3}L & \cellcolor[HTML]{F4C7C3}F1 & \cellcolor[HTML]{B7E1CD}R & 9.05 {[}3.05{] } & \cellcolor[HTML]{F4C7C3}C & \cellcolor[HTML]{F4C7C3}L & \cellcolor[HTML]{F4C7C3}F1 & \cellcolor[HTML]{F4C7C3}R  \\
    
    &Centre  & 6.32 {[}2.76{] } & \cellcolor[HTML]{F4C7C3}F & \cellcolor[HTML]{F4C7C3}L & \cellcolor[HTML]{F4C7C3}F1 & \cellcolor[HTML]{B7E1CD}R & 8.93 {[}3.02{] } & \cellcolor[HTML]{F4C7C3}F & \cellcolor[HTML]{F4C7C3}L & \cellcolor[HTML]{F4C7C3}F1 & \cellcolor[HTML]{B7E1CD}R  \\
    
    &Last & 6.50 {[}2.92{] } & \cellcolor[HTML]{F4C7C3}F & \cellcolor[HTML]{F4C7C3}C & \cellcolor[HTML]{B7E1CD}F1 & \cellcolor[HTML]{B7E1CD}R & 8.90 {[}2.96{] } & \cellcolor[HTML]{F4C7C3}F & \cellcolor[HTML]{F4C7C3}C & \cellcolor[HTML]{F4C7C3}F1 & \cellcolor[HTML]{B7E1CD}R  \\
    
    &First+1 & 6.16 {[}2.69{] } & \cellcolor[HTML]{F4C7C3}F & \cellcolor[HTML]{F4C7C3}C & \cellcolor[HTML]{B7E1CD}L  & \cellcolor[HTML]{B7E1CD}R & 8.90 {[}3.05{] } & \cellcolor[HTML]{F4C7C3}F & \cellcolor[HTML]{F4C7C3}C & \cellcolor[HTML]{F4C7C3}L  & \cellcolor[HTML]{B7E1CD}R   \\
    
    &Random & 5.86 {[}2.42{] } & \cellcolor[HTML]{B7E1CD}F & \cellcolor[HTML]{B7E1CD}C & \cellcolor[HTML]{B7E1CD}L  & \cellcolor[HTML]{B7E1CD}F1 & 9.27 {[}2.85{] } & \cellcolor[HTML]{F4C7C3}F & \cellcolor[HTML]{B7E1CD}C & \cellcolor[HTML]{B7E1CD}L  & \cellcolor[HTML]{B7E1CD}F1 \\\bottomrule
    \multicolumn{5}{c}{ }
    
    \end{tabular}
\caption{MAAE and MAE for HA, E2A, MD and FA for the different sampling scheme choices compared to using all the available repetitions. We also report the statistical significance of comparing the distributions of pairs of repetitions strategies. This provides information on whether two repetitions-sampling strategies produce the same distribution of errors across our dataset or not. In this context, a green background signifies that we can reject the null hypothesis that the errors belong to the same distribution according to the Kolmogorov–Smirnov test.}
\label{tab:bh-sampling-patterns}
\end{table}

\subsection{Deep-learning-based de-noising}
    We report the results of our experiments on tensor de-noising in Table \ref{tab:table-errors-all-models}.
    
    \subsubsection{Training additions}
    Channel normalisation brought an overall improvement compared to the baseline, especially when considering metrics computed from datasets with a higher number of repetitions. On the other hand, tensor-to-tensor training on its own appeared to be unstable, greatly benefiting some metrics while making some others worse (e.g., HA for 1BH compared to FA for 5BH) with no discernible pattern. By combining the two, we obtained a model that was more stable than one with T2T only but with slightly worse performance than using only CN. Nonetheless, T2T opened the doors to multi-tensor-to-tensor training, which brought a remarkable improvement to all metrics compared to a naive T2T approach.
    
    \subsubsection{State-of-the-art models}
    Between the two explored SOTA models, Restomer consistently outperformed Uformer at the cost of a much longer and computationally-expensive training (13 hours vs 3 hours on our machine). For this reason, the Uformer was chosen for further exploration.
    
    \subsubsection{GAN Uformer and ensemble learning}
    The addition of a discriminator and its associated loss to the training produced tensors that better encoded angular information but that encode scalar information marginally worse.
    
    The best possible model given our training additions and architectural choices was produced by an ensemble of five Wasserstein GAN Uformer models (WGUFx5). Using even a naive bagging ensemble greatly improved all metrics for all datasets compared to using a single model.
    
    All metrics except for MD for 1BH and MD for 5BH were significantly improved by our final model compared to the baseline (Wilcoxon signed-rank test, P$<$0.05). Our best performing model is the result of a naive bagging ensembling of five WGANUformer models (WGUFx5). The results show that even a naive bagging ensemble improves stability and validation performance, and reduces the variance of the output, all desirable properties in a medical setting.
    
\begin{table}[hbt]
\centering
\begin{tabular}{@{}c|l|rrrr@{}}
\toprule
\textbf{}              & \textbf{Model}                          & \textbf{HA}               & \textbf{E2A}              & \textbf{MD ($\times 10^5$)}               & \textbf{FA ($\times 10^2$)}               \\  \specialrule{.1em}{.1em}{.1em}
\multirow{12}{*}{1BH } & WGUFx5 + CN + multi T2T   & \ftextbf{11.55 {[}4.45{]}} & \ftextbf{21.79} {[}7.28{]} & 6.45 {[}3.35{]}  & 9.16 {[}4.32{]}  \\
                     & WGUF + CN + multi T2T     & 12.03 {[}4.51{]} & 22.82 {[}8.61{]} & 7.12 {[}3.20{]}  & \ftextbf{8.62 {[}3.71{]}}  \\ \cmidrule{2-6}
                     & Restomer + CN +multi T2T  & 12.07 {[}4.85{]} & 22.88 {[}7.61{]} & 6.49 {[}3.35{]}  & 9.70 {[}4.07{]}   \\
                     & Uformer + CN +multi T2T   & 12.18 {[}4.47{]} & 23.08 {[}6.27{]} & 6.52 {[}3.36{]}  & \underline{8.91 {[}4.12{]}}  \\ \cmidrule{2-6}
                     & BL + CN + multi T2T       & \underline{11.92 {[}4.30{]}} & \underline{22.17 {[}7.18{]}} & 6.44 {[}3.57{]}  & 9.26 {[}4.04{]}  \\
                     & BL + CN + T2T             & 12.73 {[}5.11{]} & 25.04 {[}8.38{]} & 6.16 {[}3.05{]}  & 10.15 {[}4.73{]} \\
                     & BL + T2T                  & 13.03 [4.74]     & 26.16 [8.10]     & \ftextbf{6.03 [3.03]}       & 10.70 [4.86] \\
                     & BL + ChanNorm             & 13.60 {[}5.05{]} & 24.80 {[}8.94{]} & 6.30 {[}6.14{]}  & 9.44 {[}3.99{]}  \\ \cmidrule{2-6}
                     & Baseline (BL)             & 13.86 {[}4.36{]} & 25.90 {[}7.69{]} & \underline{6.14 {[}2.98{]}}  & 8.93 {[}3.45{]}  \\ \cmidrule{2-6}
                     & Least squares             & 20.80 [8.06] & 31.62 [7.00] & 11.40 [3.25] & 17.06 [4.65] \\ \specialrule{.1em}{.1em}{.1em}
                     
\multirow{12}{*}{3BH } & WGUFx5 + CN + multi T2T     & \ftextbf{10.05 {[}3.86{]}} & \ftextbf{18.81 {[}6.04{]}} & \ftextbf{5.02 {[}2.11{]}}  & 7.69 {[}3.72{]}  \\
                     & WGUF + CN + multi T2T       & \underline{10.25 {[}4.21{]}} & 19.56 {[}6.33{]} & \underline{5.31 {[}2.04{]}}  & 8.06 {[}3.57{]}  \\ \cmidrule{2-6}
                     & Restomer + CN + multi T2T & 10.38 {[}4.11{]} & \underline{19.41 {[}5.82{]}} & 5.72 {[}2.04{]}  & \underline{7.48 {[}3.20{]}}   \\
                     & Uformer + CN + multi T2T   & 10.42 {[}4.18{]} & 19.86 {[}7.13{]} & 5.43 {[}2.11{]}  & \ftextbf{7.26 {[}3.36{]}}  \\ \cmidrule{2-6}
                     & BL + CN + multi T2T        & 10.73 {[}5.01{]} & 20.21 {[}6.92{]} & 5.57 {[}2.01{]}   & 8.38 {[}3.19{]}  \\
                     & BL + CN + T2T              & 11.33 {[}5.30{]}  & 22.12 {[}7.84{]} & 5.54 {[}2.75{]}  & 9.55 {[}5.17{]}  \\
                     & BL + T2T & 12.57 [5.65] & 22.46 [7.40] & 5.96 [2.53] & 10.09 [4.01]\\
                     & BL + ChanNorm              & 12.85 {[}4.90{]}  & 21.62 {[}6.86{]} & 5.37 {[}1.93{]}  & 8.57 {[}3.17{]}  \\ \cmidrule{2-6}
                     & Baseline (BL)             & 12.00 {[}5.25{]}  & 24.34 {[}6.56{]} & 5.39 {[}1.78{]}  & 8.35 {[}3.12{]}  \\ \cmidrule{2-6}
                     & Least squares                      & 15.11 [7.56] & 23.85 [6.61] & 7.88 [3.39] & 12.09 [4.07] \\ \specialrule{.1em}{.1em}{.1em}
                     
\multirow{12}{*}{5BH } & WGUFx5 + CN + multi T2T     & \ftextbf{8.39 {[}3.80{]}}   & \ftextbf{15.56 {[}6.46{]}} & 4.67 {[}1.89{]}  & \ftextbf{6.30 {[}2.85{]}}   \\
                     & WGUF + CN + multi T2T       & 8.54 {[}4.15{]}  & 16.58 {[}6.07{]} & 4.53 {[}2.00{]}   & 6.61 {[}3.11{]}  \\ \cmidrule{2-6}
                     & Restomer + CN + multi T2T & 8.66 {[}3.82{]}  & \underline{15.57 {[}6.67{]}} & 4.73 {[}1.81{]}   & 6.79 {[}2.80{]}   \\
                     & Uformer + CN + multi T2T   & 8.75 {[}4.03{]}  & 16.07 {[}6.43{]} & 4.48 {[}1.75{]}  & \underline{6.44 {[}2.48{]}}  \\ \cmidrule{2-6}
                     & BL + CN + multi T2T        & \underline{8.45 {[}4.25{]}}  & 16.30 {[}6.99{]}  & 4.79 {[}2.03{]}  & 6.71 {[}2.79{]}  \\
                     & BL + CN + T2T              & 9.54 {[}5.44{]}  & 17.20 {[}7.34{]}  & 5.71 {[}2.69{]}  & 9.23 {[}3.91{]}  \\
                     & BL + T2T              & 9.77 [4.90] & 17.77 [7.15] & 4.59 [1.80] & 7.78 [3.12]\\
                     & BL + ChanNorm              & 10.81 {[}4.11{]} & 18.92 {[}6.60{]}  & \underline{4.16 {[}1.74{]}}  & 6.96 {[}2.43{]}  \\ \cmidrule{2-6}
                     & Baseline (BL)             & 10.94 {[}3.80{]}  & 19.61 {[}5.37{]} & \ftextbf{4.11 {[}2.14{]}}  & 8.03 {[}3.34{]}  \\ \cmidrule{2-6}
                     & Least squares                      & 9.79 [4.26] & 17.20 [5.51] & 6.10 [2.67] & 8.30 [2.93]  \\ \bottomrule
\multicolumn{5}{c}{ }
\end{tabular}
\caption{DT-CMR maps errors for all the deep-learning models we experiment with. The table is divided into three sections, one per dataset (1BH, 3BH, 5BH) and also include the error for the linear approximation method of the maps (Least squares). In bold and underlined we report, respectively, the best and second best results for each metric and dataset.}
\label{tab:table-errors-all-models}
\end{table}

\section{Discussion}
    
    \subsection{Breath-hold choice}
    From our results on the breath-hold repetition sampling patterns, we can draw several conclusions on how the patient behaviour affects the quality of the maps:
    \begin{itemize}
        \item Despite the clinician's intuition, no significant conclusion can be drawn about the difference in errors between the First and the First+1 protocols.
        \item There is no clear pattern for the pairwise significance of the results within a dataset, suggesting that there are other factors that affect the quality of the maps besides the sampling pattern.
        \item Lower numbers of breath-holds produce far fewer significant results compared to higher-breath-holds datasets (44/80 for BH1, 46/80 for BH3 and 48/80 for BH5). This can be attributed to the higher variability of errors due to the effect of the singular bad acquisition on the final quality of the maps. Such effect is instead smoothed out when considering the maps produced from a larger number of repetitions, making the error distributions more similar to each other.
    \end{itemize}
    Keeping the statistical significance in mind, First, Centre, or Last sampling strategies seem a reasonable choice as they have lower error and no significant difference in error distributions. The exception is for MD, for which the difference is sometimes significant. For this reason, we decided to train our models on the First, Centre, and Last strategies, but to only use the First strategy for the validation and test sets. Notice that the data acquired with the First+1 strategy was not used for the training as it contains a considerable amount of redundancy with that acquired with the First scheme. This also mirrors a real clinical acquisition situation, where choosing the first M repetitions is the shortest option in terms of the number of breath-holds for the patient. Any other strategy would require us to acquire and discard some data, which is not feasible in a clinical setting where the aim is to minimise the scan time for the patient.
    
    \subsection{Deep-learning-based de-noising}
    Our proposed additions to the training (channel normalisation, tensor-to-tensor training and multi-tensor-to-tensor training) have a beneficial effect if we consider the errors on the DT-CMR derived maps. This is due to various reasons:
    \begin{itemize}
        \item Channel normalisation simplifies the training, allowing the network to not focus on rescaling the output to match the input range for each channel individually. 
        \item Tensor-to-tensor training completely changes the training objective. In our previous work, we had trained a model to compute de-noised tensors from noisy images, effectively replacing the linear optimisation problem (LSS). This corresponds to training the model on two tasks simultaneously: de-noising and tensor computation, making the overall convergence harder. In our new proposed setting, we simplify the training objective by removing the tensor-computation aspect and only leaving the de-noising part of the training. Moreover, this allowed us to make use of the existing literature in the well-studied field of image de-noising, while our previous approach (image-to-tensor) proposed a model for a completely novel task with no existing literature.
        \item Making use of multiple sampling patterns from our available data also gave us an advantage over previous work by allowing us to greatly increase the size of the dataset used for the training without the need to scan additional patients. 
    \end{itemize}
    Our final model, WGUFx5, draws from the SOTA in camera images de-noising uses several novel blocks to encode local information by using local self-attention and hierarchical feature encoding. All these additions produce a definite improvement in our tensor de-noising task.
    
    Finally, according to the literature, a bagging ensemble reduces the variance of the prediction and therefore suggests that previous models were inadvertently overfitting to the training set, despite our efforts to prevent it. The ensemble strategy in our setting can be therefore interpreted to act as a regulariser.

\section{Conclusion}
    DT-CMR has the potential to revolutionise the ability to non-invasively image and assess the microstructural organisation of the myocardium underlying cardiac pathology, but it is held back from clinical translation by its long acquisition times. Here, we proposed to tackle the problem by reducing the number of repetitions used in a classical DT-CMR acquisition protocol, which linearly reduced the total acquisition time, but also decreased SNR. When choosing the repetition selection scheme, we demonstrated that the choice of breath-hold had no statistically significant effect on the final quality of the DT-CMR maps.  
    We also proposed several improvements on existing deep learning models, that, combined, may lead to a significant and considerable step towards single-breath-hold DT-CMR acquisition for clinical use.

\clearpage
\bibliographystyle{splncs04}
\bibliography{PhD,manual}

\begin{thebibliography}{10}
\providecommand{\url}[1]{\texttt{#1}}
\providecommand{\urlprefix}{URL }
\providecommand{\doi}[1]{https://doi.org/#1}

\bibitem{arjovskyWassersteinGAN2017}
Arjovsky, M., Chintala, S., Bottou, L.: Wasserstein {{GAN}}. arXiv:1701.07875
  [cs, stat]  (Dec 2017)

\bibitem{basser1995inferring}
Basser, P.J.: Inferring microstructural features and the physiological state of
  tissues from diffusion-weighted images. NMR in Biomedicine  \textbf{8}(7),
  333--344 (1995)

\bibitem{batsonNoise2SelfBlindDenoising2019}
Batson, J., Royer, L.: {{Noise2Self}}: {{Blind Denoising}} by
  {{Self-Supervision}}. In: Proceedings of the 36th {{International
  Conference}} on {{Machine Learning}}. pp. 524--533. {PMLR} (May 2019)

\bibitem{chenPreTrainedImageProcessing2021}
Chen, H., Wang, Y., Guo, T., Xu, C., Deng, Y., Liu, Z., Ma, S., Xu, C., Xu, C.,
  Gao, W.: Pre-{{Trained Image Processing Transformer}}. In: 2021
  {{IEEE}}/{{CVF Conference}} on {{Computer Vision}} and {{Pattern
  Recognition}} ({{CVPR}}). pp. 12294--12305. {IEEE}, {Nashville, TN, USA} (Jun
  2021). \doi{10.1109/CVPR46437.2021.01212}

\bibitem{chenLowdoseCTDenoising2017}
Chen, H., Zhang, Y., Zhang, W., Liao, P., Li, K., Zhou, J., Wang, G.: Low-dose
  {{CT}} denoising with convolutional neural network. In: 2017 {{IEEE}} 14th
  {{International Symposium}} on {{Biomedical Imaging}} ({{ISBI}} 2017). pp.
  143--146 (Apr 2017). \doi{10.1109/ISBI.2017.7950488}

\bibitem{IterativeReweightedLinear}
Collier, Q., Veraart, J., Jeurissen, B., den Dekker, A.J., Sijbers, J.:
  {Iterative reweighted linear least squares for accurate, fast, and robust
  estimation of diffusion magnetic resonance parameters} (2015)

\bibitem{demirPatchBasedImageInpainting2018}
Demir, U., Unal, G.: Patch-{{Based Image Inpainting}} with {{Generative
  Adversarial Networks}}. arXiv:1803.07422 [cs]  (Mar 2018)

\bibitem{guizar-sicairosEfficientSubpixelImage2008}
{Guizar-Sicairos}, M., Thurman, S.T., Fienup, J.R.: Efficient subpixel image
  registration algorithms. Optics Letters  \textbf{33}(2),  156--158 (Jan
  2008). \doi{10.1364/OL.33.000156}

\bibitem{hermosillaTotalDenoisingUnsupervised2019}
Hermosilla, P., Ritschel, T., Ropinski, T.: Total {{Denoising}}: {{Unsupervised
  Learning}} of {{3D Point Cloud Cleaning}}. In: Proceedings of the
  {{IEEE}}/{{CVF International Conference}} on {{Computer Vision}}. pp. 52--60
  (2019)

\bibitem{jiangDenoising3DMagnetic2018}
Jiang, D., Dou, W., Vosters, L., Xu, X., Sun, Y., Tan, T.: Denoising of {{3D}}
  magnetic resonance images with multi-channel residual learning of
  convolutional neural network. Japanese Journal of Radiology  \textbf{36}(9),
  566--574 (Sep 2018). \doi{10.1007/s11604-018-0758-8}

\bibitem{IntroductionDiffusionTensor}
Kingsley, P.B.: {Introduction to diffusion tensor imaging mathematics: Part
  III. Tensor calculation, noise, simulations, and optimization} (2006)

\bibitem{kung2011presence}
Kung, G.L., Nguyen, T.C., Itoh, A., Skare, S., Ingels~Jr, N.B., Miller, D.C.,
  Ennis, D.B.: The presence of two local myocardial sheet populations confirmed
  by diffusion tensor mri and histological validation. Journal of Magnetic
  Resonance Imaging  \textbf{34}(5),  1080--1091 (2011)

\bibitem{liangSwinIRImageRestoration2021}
Liang, J., Cao, J., Sun, G., Zhang, K., Van~Gool, L., Timofte, R.: {{SwinIR}}:
  {{Image Restoration Using Swin Transformer}}. In: 2021 {{IEEE}}/{{CVF
  International Conference}} on {{Computer Vision Workshops}} ({{ICCVW}}). pp.
  1833--1844. {IEEE}, {Montreal, BC, Canada} (Oct 2021).
  \doi{10.1109/ICCVW54120.2021.00210}

\bibitem{loshchilovDecoupledWeightDecay2019}
Loshchilov, I., Hutter, F.: Decoupled {{Weight Decay Regularization}}.
  arXiv:1711.05101 [cs, math]  (Jan 2019)

\bibitem{manjonMRIDenoisingUsing2018}
Manj{\'o}n, J.V., Coupe, P.: {{MRI Denoising Using Deep Learning}}. In: Bai,
  W., Sanroma, G., Wu, G., Munsell, B.C., Zhan, Y., Coup{\'e}, P. (eds.)
  Patch-{{Based Techniques}} in {{Medical Imaging}}, vol. 11075, pp. 12--19.
  {Springer International Publishing}, {Cham} (2018).
  \doi{10.1007/978-3-030-00500-9}

\bibitem{moriPrinciplesDiffusionTensor2006}
Mori, S., Zhang, J.: Principles of {{Diffusion Tensor Imaging}} and {{Its
  Applications}} to {{Basic Neuroscience Research}}. Neuron  \textbf{51}(5),
  527--539 (Sep 2006). \doi{10.1016/j.neuron.2006.08.012}

\bibitem{parkUnpairedImageDenoising2019}
Park, H.S., Baek, J., You, S.K., Choi, J.K., Seo, J.K.: Unpaired {{Image
  Denoising Using}} a {{Generative Adversarial Network}} in {{X-Ray CT}}. IEEE
  Access  \textbf{7},  110414--110425 (2019). \doi{10.1109/ACCESS.2019.2934178}

\bibitem{phippsAcceleratedVivoCardiac2021}
Phipps, K., {van de Boomen}, M., Eder, R., Michelhaugh, S.A., Spahillari, A.,
  Kim, J., Parajuli, S., Reese, T.G., Mekkaoui, C., Das, S., Gee, D., Shah, R.,
  Sosnovik, D.E., Nguyen, C.: Accelerated in {{Vivo Cardiac Diffusion-Tensor
  MRI Using Residual Deep}} {{Learning}}\textendash based {{Denoising}} in
  {{Participants}} with {{Obesity}}. Radiology: Cardiothoracic Imaging
  \textbf{3}(3),  e200580 (Jun 2021). \doi{10.1148/ryct.2021200580}

\bibitem{schlaudeckerGadoliniumAssociatedNephrogenicSystemic2009}
Schlaudecker, J.D., Bernheisel, C.R.: Gadolinium-{{Associated Nephrogenic
  Systemic Fibrosis}}. American Family Physician  \textbf{80}(7),  711--714
  (Oct 2009)

\bibitem{wangUformerGeneralUShaped2021}
Wang, Z., Cun, X., Bao, J., Zhou, W., Liu, J., Li, H.: Uformer: {{A General
  U-Shaped Transformer}} for {{Image Restoration}}. arXiv:2106.03106 [cs]  (Nov
  2021)

\bibitem{OURPAPERNOTPUBLIC}
X, X.: This work has been accepted to jmri 2022 but it is not publicly
  available. all the information has been anonymised. In: X. {JMRI} (2022)

\bibitem{OURPAPERNOTPUBLIC2}
X, X.: Previous work by our research group. {A}ll the information has been
  anonymised. In: X. {X} (20XX)

\bibitem{xieImageDenoisingInpainting2012}
Xie, J., Xu, L., Enhong, C.: Image {{Denoising}} and {{Inpainting}} with {{Deep
  Neural Networks}}. In: Advances in {{Neural Information Processing Systems}}.
  vol.~3, pp. 183--189. {Morgan-Kaufmann} (2012)

\bibitem{zamirRestormerEfficientTransformer2021}
Zamir, S.W., Arora, A., Khan, S., Hayat, M., Khan, F.S., Yang, M.H.: Restormer:
  {{Efficient Transformer}} for {{High-Resolution Image Restoration}}.
  arXiv:2111.09881 [cs]  (Nov 2021)

\bibitem{zhangGaussianDenoiserResidual2017}
Zhang, K., Zuo, W., Chen, Y., Meng, D., Zhang, L.: Beyond a {{Gaussian
  Denoiser}}: {{Residual Learning}} of {{Deep CNN}} for {{Image Denoising}}.
  IEEE Transactions on Image Processing  \textbf{26}(7),  3142--3155 (Jul
  2017). \doi{10.1109/TIP.2017.2662206}

\end{thebibliography}

\clearpage
\appendix

    




\end{document}